\documentclass[10pt,prc,aps,superscriptaddress]{revtex4}   
\usepackage{graphicx}
\tighten
\begin{document}
\draft
\title{                                                     
 Dirac-Brueckner-Hartree-Fock {\it versus} chiral effective
 field theory                           
 }              
\author{            
 Francesca Sammarruca}                                                                                
\affiliation{                 
Physics Department, University of Idaho, Moscow, ID 83844-0903, U.S.A  } 
\author{B. Chen}    
\affiliation{                 
Physics Department, University of Idaho, Moscow, ID 83844-0903, U.S.A  } 
\author{L. Coraggio} 
\affiliation{                 
Istituto Nazionale di Fisica Nucleare,  \\ 
Complesso Universitario di Monte S. Angelo, Via Cintia - I-80126 Napoli, Italy } 
\author{N. Itaco} 
\affiliation{                 
Departimento di Scienze Fisiche, Universit\'{a} di Napoli Federico II, 
Complesso Universitario di Monte S. Angelo, Via Cintia - I-80126 Napoli, Italy } 
\affiliation{                 
Istituto Nazionale di Fisica Nucleare,  \\ 
Complesso Universitario di Monte S. Angelo, Via Cintia - I-80126 Napoli, Italy } 
\author{R. Machleidt} 
\affiliation{                 
Physics Department, University of Idaho, Moscow, ID 83844-0903, U.S.A  } 
\date{\today} 
\begin{abstract}
We compare nuclear and neutron matter predictions based on two different 
{\it ab initio} approaches to nuclear forces and the nuclear many-body
problem. The first consists of a realistic meson-theoretic nucleon-nucleon
potential together with the relativistic counterpart of the Brueckner-Hartree-Fock 
theory of nuclear matter. The second is based on chiral effective field theory,
with density-dependent interactions derived from leading order chiral three-nucleon
forces. We find the results to be very close and conclude that both approaches 
contain important features governing the physics of nuclear and neutron matter.           
\end{abstract}
\maketitle

\section{Introduction} 
                                                                     
Nuclear matter is a convenient laboratory to test nuclear forces and many body theories. 
In particular, the equation of state (EoS) of 
extremely neutron-rich matter has attracted considerable attention lately because of its broad applications,             
ranging from the structure of 
rare isotopes to the properties of neutron stars.

Constraints on the properties of nuclear matter, symmetric or isospin-asymmetric, can be obtained
from a variety of experiments, such as measurements of 
nuclear binding energies (including isobaric analog state energies), parity-violating electron
scattering, neutron skin thickness measurements, nucleus-nucleus collisions, and astrophysical 
observations. For a recent review of available constraints, particularly on the nuclear symmetry energy, the reader is referred to Ref.~\cite{Tsang+}. 

Diverse 
theoretical frameworks have been employed to describe the properties of nuclear and neutron matter. They
include: phenomenological approaches, both relativistic and non-relativistic;         
the Brueckner-Hartree-Fock (BHF) method, typically implemented with three-nucleon forces (3NF); 
variational approaches; the relativistic 
Dirac-Brueckner-Hartree-Fock (DBHF) approach; and chiral effective field theories. Predictions are
model-dependent, particularly at the higher densities, where constraints are scarse and less 
stringent.

\begin{figure}
\centering            
\vspace*{-3.2cm}
\hspace*{0.5cm}
\scalebox{1.0}{\includegraphics{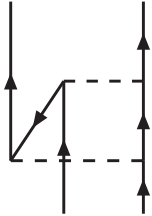}}
\vspace*{-21.0cm}
\caption{Contribution to the NN interaction from virtual pair excitation.                   
Upward- and downward-pointing arrows represent nucleons and antinucleons, respectively.
Dashed lines denote mesons.                            
} 
\label{2b}
\end{figure}

In our previous work with both symmetric nuclear matter (SNM) and isospin-asymmetric 
nuclear matter (IANM), we have relied on the DBHF approach, together with a relativistic 
meson-theoretic nucleon-nucleon (NN) potential which uses the pseudovector coupling for the pion 
\cite{Sam10}. 
The latter choice was motivated by the considerations we outline next. 
Already when QCD (and its approximate symmetries) were unknown, it was observed that the contribution from the
nucleon-antinucleon pair diagram, Fig.~\ref{2b}, becomes unreasonably large if the pseudoscalar (ps) coupling is used 
for the pion,
leading to very large pion-nucleon scattering lengths \cite{Wei68,GB79}.                                            
We recall that the Lagrangian density for pseudoscalar coupling of the nucleon field ($\psi$) with a pseudoscalar meson
field ($\phi$) is 
\begin{equation}
{\cal L}_{ps} = -ig_{ps}\bar {\psi} \gamma _5 \psi \phi.     \label{ps} 
\end{equation} 
On the other hand, the same contribution shown in Fig.~\ref{2b}, 
is heavily suppressed when the pseudovector (pv) coupling is used instead (a mechanism which
became known as ``pair suppression"). The reason for the suppression is that the             
covariant derivative                                                                                     
in the  pseudovector Lagrangian,                                              
\begin{equation}
{\cal L}_{pv} = -\frac{f_{ps}}{m_{ps}}{\bar \psi}  \gamma _5 \gamma^{\mu}\psi \partial_{\mu} \phi \; ,          
\label{pv} 
\end{equation} 
generates a vertex that is proportional to momentum (leading to a weak coupling for low momenta) and, thus, 
 explains the small value of the pion-nucleon
scattering length at threshold.               
Non-linear realizations of chiral symmetry \cite{Wei68} can further motivate a preference for 
the pseudovector coupling.                          

\begin{figure}[!t] 
\centering         
\vspace*{-3.2cm}
\hspace*{0.5cm}
\scalebox{0.9}{\includegraphics{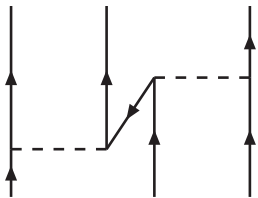}}
\vspace*{-19.0cm}
\caption{Three-body force due to virtual pair excitation. Notation as in the previous figure.
} 
\label{3b}
\end{figure}

Back to the many-body problem,   
the main strength of the DBHF approach is its inherent ability to account for important three-body forces   
through its density dependence. 
In Fig.~\ref{3b} we show a three-body force originating from virtual excitation of a nucleon-antinucleon pair, 
known as ``Z-graph". Notice that the observations from the previous paragraph ensure that this diagram, as well as the corresponding diagram
at the two-body level, Fig.~\ref{2b}, is moderate in size when the pv coupling,           
Eq.~(\ref{pv}), is used. (Hence, the importance of using pv potentials in relativistic nuclear structure
calculations.)  
The main feature of                                     
the DBHF method turns out to be closely related to 
the 3NF depicted in Fig.~\ref{3b}, as we will argue next. In the DBHF approach, one describes the positive energy solutions
of the Dirac equation in the medium as 
\begin{equation}
u^*(p,\lambda) = \left (\frac{E^*_p+m^*}{2m^*}\right )^{1/2}
\left( \begin{array}{c}
 {\bf 1} \\
\frac{\sigma \cdot \vec {p}}{E^*_p+m^*} 
\end{array} 
\right) \;
\chi_{\lambda},
\label{ustar}
\end{equation}
where the nucleon effective mass, $m^*$, is defined as $m^* = m+U_S$, with $U_S$ an attractive scalar potential.
(This will be derived below.) 
It can be shown that both the description of a single-nucleon via Eq.~(\ref{ustar}) and the evaluation of the 
Z-diagram, Fig.~\ref{3b}, generate a repulsive effect on the energy/particle in symmetric nuclear matter which depends on the density approximately
as 
\begin{equation}
\Delta E \propto  \left (\frac{\rho}{\rho_0}\right )^{8/3} \, , 
\label{delE} 
\end {equation}
and provides the saturating mechanism missing from conventional Brueckner calculations.

The approximate equivalence of the effective-mass description of Dirac states and the contribution from the Z-diagram 
has a simple intuitive explanation in the observation 
that Eq.~(\ref{ustar}), like any other solution of the Dirac equation,
can be written as a superposition of positive and negative energy solutions using free nucleon masses. On the other hand, the ``nucleon" in the 
middle of the Z-diagram, Fig.~\ref{3b}, is precisely a superposition of positive and negative energy states. 
In summary, the DBHF method effectively takes into account a particular class of 
3NF, which are crucial for nuclear matter saturation. 

The conventional BHF theory, together with meson-exchange 3NF, can also describe the saturation of
nuclear matter in a satisfactory manner, although consistency between the parameters of the 
two- and the three-body forces can be problematic \cite{Catania}.

Ideally, one wishes to derive nuclear forces from the fundamental theory of strong interactions, QCD.
Such task is however not feasible, due to the   
non-perturbative nature of the theory in the low-energy regime typical of nuclear physics. 
An alternative way is to respect the 
symmetries of the QCD Lagrangian while retaining the traditional degrees of freedom relevant
to nuclear physics, namely nucleons and pions \cite{Wein79}. 
This is the philosophy of chiral 
effective field theorie (EFT), which has become popular in recent years.                 

Effective field theories 
 allow for a systematic expansion in powers of the momentum (or the pion mass) known as chiral perurbation theory (ChPT) such that, at any order,
the irreducible two- and many-body diagrams to be included are precisely defined.
An extensive review on the subject, including a comprehensive list or references, can be found in Ref.~\cite{ME11}.     

Chiral EFT has validity only up to the chiral symmetry breaking scale of $\Lambda_{\chi}\approx 1$ GeV.
Thus, 
the low-momentum expansion has a limited range of applicability. 
This is another reason why relativistic meson theory has been considered more suitable
for applications to dense systems, where high momenta are involved due to the high Fermi
momenta.   
To remedy this problem, ways to extend EFT predictions to higher densities are employed,
such as parametrizing the available predictions with accurate fitting functions which are then used to         
predict the properties of dense astrophysical systems.

It is the purpose of this paper to conduct a comparison between nuclear and neutron matter (NM) 
predictions obtained with two {\it ab-initio} approaches:
\begin{enumerate}
\item 
 The DBHF method as briefly outlined above, using the Bonn B meson-exchange potential \cite{Mac89}. 
\item 
 A quantitative chiral NN potential, for which we choose the Idaho N$^3$LO potential of Ref.~\cite{EM03} 
together with chiral three-nucleon forces at N$^2$LO. Details are given in the next section. 
\end{enumerate}

We would like to explore how the DBHF phenomenology compares with 
the predictions of chiral EFT 
 at N$^3$LO plus leading 3NF, in both symmetric and neutron matter. 

We will start with a comparison at the two-body force level. The meson-exchange side
of this part of the comparison will be represented by the conventional BHF model.
In the meson-exchange sector, realistic saturation properties will be generated through 
the DBHF model, whereas, on the chiral side, the contribution from leading 3NF will be included. 
We wish to compare the size and density dependence of the saturating effects in each
scheme. We will compare predictions of the equation of state and the symmetry energy in both cases.
Neutron star masses and radii will also be addressed. 

This paper is organized as follows:                                                  
In Section II, we review the main points of the DBHF calculation leading to the energy per
particle in nuclear matter with arbitrary degree of isospin asymmetry.
A brief outline of chiral two- and three-nucleon forces is contained in Section III. 
Our findings and conclusions are presented in Sections IV and V, respectively.

\section{Relativistic meson-exchange potential and DBHF approach} 

As stated in the Introduction, 
the starting point of our many-body calculation is a realistic NN interaction which is then applied in the 
nuclear medium without any additional free parameters. 

Relativistic meson theory is an appropriate framework to deal with the high momenta encountered in dense
matter. In particular, 
the one-boson-exchange (OBE) model has proven very successful in describing NN data in free space 
up to high energy and has a good theoretical foundation. 
The OBE potential is defined as a sum of one-particle-exchange amplitudes of certain bosons with 
given mass and coupling. In general, six non-strange bosons with masses below 1 GeV$/c^2$ are used.
Thus,
\begin{equation}
v = \sum_{\alpha=\pi,\eta,\rho,\omega,\delta,\sigma} v_{\alpha}^{OBE}  \;,
\end{equation}
with $\pi$ and $\eta$ pseudoscalar, $\sigma$ and $\delta$ scalar, and $\rho$ and $\omega$ vector
particles. For more details, see Ref.~\cite{Mac89}.   

Among the many available OBE potentials, some being part of the ``high-precision generation" \cite{pot1,pot2}, 
we seek a momentum-space potential developed within a relativistic scattering equation, such as the 
one obtained through the Thompson \cite{Thom} three-dimensional reduction of the Bethe-Salpeter equation \cite{BS}. 

Having summarized  in the Introduction the main DBHF philosophy, for completeness we
 now proceed to review the main aspects of our approach and the various approximations 
we perform through the application of the DBHF procedure. 
The equations we present are those suitable for isospin-asymmetric nuclear matter (IANM), since they
naturally 
contain both the cases of SNM and NM. 

We start from the Thompson \cite{Thom} relativistic three-dimensional reduction 
of the Bethe-Salpeter equation \cite{BS}. The Thompson equation is applied to nuclear matter in
strict analogy to free-space scattering and reads, in the nuclear matter rest frame,                 
\begin{eqnarray}
&& g_{ij}(\vec q',\vec q,\vec P,(\epsilon ^*_{ij})_0) = v_{ij}^*(\vec q',\vec q) \nonumber \\            
&& + \int \frac{d^3K}{(2\pi)^3}v^*_{ij}(\vec q',\vec K)\frac{m^*_i m^*_j}{E^*_i E^*_j}
\frac{Q_{ij}(\vec K,\vec P)}{(\epsilon ^*_{ij})_0 -\epsilon ^*_{ij}(\vec P,\vec K)} 
g_{ij}(\vec K,\vec q,\vec P,(\epsilon^*_{ij})_0) \, ,                                   
\label{gij}
\end{eqnarray}                    
where $g_{ij}$ is the in-medium reaction matrix 
($ij$=$nn$, $pp$, or $np$), and the                                      
asterix signifies that medium effects are applied to those quantities. Thus the NN potential, 
$v_{ij}^*$, is constructed in terms of effective Dirac states (in-medium spinors) as explained above. 
In Eq.~(\ref{gij}),                                  
$\vec q$, $\vec q'$, and $\vec K$ are the initial, final, and intermediate
relative momenta, and $E^*_i = \sqrt{(m^*_i)^2 + K^2}$. 
The momenta of the two interacting particles in the nuclear matter rest frame have been expressed in terms of their
relative momentum and the center-of-mass momentum, $\vec P$, through
\begin{equation} 
\vec P = \vec k_{1} + \vec k_{2}       \label{P}    
\end{equation} 
and 
\begin{equation} 
\vec K = \frac{\vec k_{1} - \vec k_{2}}{2} \, .  \label{K}
\end{equation}                    
The energy of the two-particle system is 
\begin{equation} 
\epsilon ^*_{ij}(\vec P, \vec K) = 
e^*_{i}(\vec P, \vec K)+  
e^*_{j}(\vec P, \vec K)   
\label{eij}
\end{equation} 
 and $(\epsilon ^*_{ij})_0$ is the starting energy.
 The single-particle energy $e_i^*$ includes kinetic energy and potential 
 energy contributions (see Eq.~(\ref{spe}) below).                               
The Pauli operator, $Q_{ij}$, prevents scattering to occupied $nn$, $pp$, or $np$ states.            
 To eliminate the angular
dependence from the kernel of Eq.~(\ref{gij}), it is customary to replace the exact
Pauli operator with its angle-average. 
Detailed expressions for the Pauli operator                     
and the average center-of-mass momentum in the case of two different Fermi seas  
can be found in Ref.~\cite{AS03}.                              

With the definitions
\begin{equation} 
G_{ij} = \frac{m^*_i}{E_i^*(\vec{q'})}g_{ij}
 \frac{m^*_j}{E_j^*(\vec{q})}             
\label{Gij}
\end{equation} 
and 
\begin{equation} 
V_{ij}^* = \frac{m^*_i}{E_i^*(\vec{q'})}v_{ij}^*
 \frac{m^*_j}{E_j^*(\vec{q})} \, ,        
\label{Vij}
\end{equation} 
 one can rewrite Eq.~(\ref{gij}) as
\begin{eqnarray}
&& G_{ij}(\vec q',\vec q,\vec P,(\epsilon ^*_{ij})_0) = V_{ij}^*(\vec q',\vec q) \nonumber \\[4pt]
&& + \int \frac{d^3K}{(2\pi)^3}V^*_{ij}(\vec q',\vec K)
\frac{Q_{ij}(\vec K,\vec P)}{(\epsilon ^*_{ij})_0 -\epsilon ^*_{ij}(\vec P,\vec K)} 
G_{ij}(\vec K,\vec q,\vec P,(\epsilon^*_{ij})_0) \, ,                                    
\label{Geq}
\end{eqnarray}                    
which is formally identical to its non-relativistic counterpart.

The goal is to determine self-consistently the nuclear matter single-particle potential   
which, in IANM, will be different for neutrons and protons. 
To facilitate the description of the procedure, we will use a schematic
notation for the neutron/proton potential.                                                   
We write, for neutrons,
\begin{equation}
U_n = U_{np} + U_{nn} \; , 
\label{un}
\end{equation}
and for protons
\begin{equation}
U_p = U_{pn} + U_{pp} \, , 
\label{up}
\end{equation}
where each of the four pieces on the right-hand-side of Eqs.~(\ref{un}-\ref{up}) signifies an integral of the appropriate 
$G$-matrix elements ($nn$, $pp$, or $np$) obtained from Eq.~(\ref{Geq}).                                           
Clearly, the two equations above are coupled through 
the $np$ component and so they must be solved simultaneously. Furthermore, 
the $G$-matrix equation and Eqs.~(\ref{un}-\ref{up})  
are coupled through the single-particle energy (which includes the single-particle
potential, itself defined in terms of the $G$-matrix). So we have a coupled system to be solved self-consistently.

Before proceeding with the self-consistency, 
one needs an {\it ansatz} for the single-particle potential. The latter is suggested by 
the most general structure of the nucleon self-energy operator consistent with 
all symmetry requirements. That is: 
\begin{equation}
{\cal U}_i({\vec p}) =  U_{S,i}(p) + \gamma_0  U_{V,i}^{0}(p) - {\vec \gamma}\cdot {\vec p}  U_{V,i}(p) \, , 
\label{Ui1}
\end{equation}
where $U_{S,i}$ and 
$U_{V,i}$ are an attractive scalar field and a repulsive vector field, respectively, with 
$ U_{V,i}^{0}$ the timelike component of the vector field. These fields are in general density and momentum dependent. 
We take             
\begin{equation}
{\cal U}_i({\vec p}) \approx U_{S,i}(p) + \gamma_0 U_{V,i}^{0}(p) \, ,                                            
\label{Ui2}
\end{equation}
which amounts to assuming that the spacelike component of the vector field is much smaller than 
 both $U_{S,i}$ and $U_{V,i}^0$. Furthermore, neglecting the momentum dependence of the scalar and
vector fields and inserting Eq.~(\ref{Ui2}) in the Dirac equation for neutrons/protons propagating in 
nuclear matter,
\begin{equation}
(\gamma _{\mu}p^{\mu} - m_i - {\cal U}_i({\vec p})) u_i({\vec p},\lambda) = 0  \, ,                                                       
\label{Dirac1} 
\end{equation}
naturally leads to rewriting the Dirac equation in the form 
\begin{equation}
(\gamma _{\mu}(p^{\mu})^* - m_i^*) u_i({\vec p},\lambda) = 0  \, ,                                                       
\label{Dirac2} 
\end{equation}
with positive energy solutions as in Eq.~(\ref{ustar}), $m_i^* = m + U_{S,i}$, and 
\begin{equation}
(p^0)^* = p^0 - U_{V,i}^0 (p) \, .                                                                 
\label{p0}
\end{equation}
The subscript ``$i$'' signifies that these parameters are different for protons and
neutrons. 

As in the symmetric matter case \cite{BM84}, evaluating  the expectation value of Eq.~(\ref{Ui2})       
leads to a parametrization of 
the single particle potential for protons and neutrons (Eqs.(\ref{un}-\ref{up})) in terms of the 
constants $U_{S,i}$ and $U_{V,i}^0$ which is given by      
\begin{equation}
U_i(p) = \frac{m^*_i}{E^*_i}<{\vec p}|{\cal U}_i({\vec p})|{\vec p}> = 
\frac{m^*_i}{E^*_i}U_{S,i} + U_{V,i}^0 \; .      
\label{Ui3}
\end{equation}
Also, 
\begin{equation}
U_i(p) =                                                              
\sum_{j=n,p} 
\sum_{p' \le k_F^j} G_{ij}({\vec p},{\vec p}') \; , 
\label{Ui4}
\end{equation}
which, along with Eq.~(\ref{Ui3}), allows the self-consistent determination of the single-particle
potential as explained below. 

The kinetic contribution to the single-particle energy is
\begin{equation}
T_i(p) = \frac{m^*_i}{E^*_i}<{\vec p}|{\vec \gamma} \cdot {\vec p} + m|{\vec p}> =     
\frac{m_i m^*_i + {\vec p}^2}{E^*_i} \; , 
\label{KE}    
\end{equation}
and the single-particle energy is 
\begin{equation}
e^*_i(p) = T_i(p) + U_i(p) = E^*_i + U^0_{V,i} \; . 
\label{spe}
\end{equation}
The constants $m_i^*$ and 
\begin{equation}
U_{0,i} = U_{S,i} + U_{V,i}^0      
\label{U0i} 
\end{equation}
are convenient to work with as they 
facilitate          
the connection with the usual non-relativistic framework \cite{HT70}.                       

Starting from some initial values of $m^*_i$ and $U_{0,i}$, the $G$-matrix equation is 
 solved and a first approximation for $U_{i}(p)$ is obtained by integrating the $G$-matrix 
over the appropriate Fermi sea, see Eq.~(\ref{Ui4}). This solution is 
again parametrized in terms of a new set of constants, determined by fitting the parametrized $U_i$, 
Eq.~(\ref{Ui3}), 
to its values calculated at two momenta, a procedure known as the ``reference spectrum approximation". 
The iterative procedure is repeated until satisfactory convergence is reached.     

Finally, the energy per neutron or proton in nuclear matter is calculated from 
the average values of the kinetic and potential energies as 
\begin{equation}
\bar{e}_{i} = \frac{1}{A}<T_{i}> + \frac{1}{2A}<U_{i}> -m \; . 
\label{ei}
\end{equation}
 The EoS, or energy per nucleon as a function of density, is then written as
\begin{equation}
    \bar{e}(\rho_n,\rho_p) = \frac{\rho_n \bar{e}_n + \rho_p \bar{e}_p}{\rho} \, , 
\label{enp} 
\end{equation}
or 
\begin{equation}
    \bar{e}(k_F,\alpha) = \frac{(1 + \alpha) \bar{e}_n + (1-\alpha) \bar{e}_p}{2} \, . 
\label{eav} 
\end{equation}
Clearly, symmetric nuclear matter is obtained as a by-product of the calculation described above 
by setting $\alpha$=0, whereas $\alpha$=1 corresponds to pure neutron matter.

\section{Chiral two- and three-nucleon forces} 

Ideally, one wishes to base a derivation of the nuclear force on QCD. However, the well-known 
problem with QCD is that it is non-perturbative in the low-energy regime characteristic for 
nuclear physics. For many years this fact was perceived as a great obstacle to a derivation of
nuclear forces from QCD--impossible to overcome except with lattice QCD. The effective field theory
concept has shown the way out of this dilemma. One has to realize that the scenario of low-energy QCD
is characterized by pions and nucleons interacting via a force governed by spontaneously broken
approximate chiral symmetry. This chiral EFT allows for a systematic low-momentum expansion known as
chiral perturbation theory (ChPT). Contributions are analyzed in terms of powers of small external
momenta over the large scale, $(Q/\Lambda_{\chi})^{\nu}$, where $Q$ is generic for an external 
momentum (nucleon three-momentum or pion four-momentum) or pion mass, and 
$\Lambda_{\chi}\approx $ 1 GeV is the chiral symmetry breaking scale (`hard scale'). 
(See Ref.~\cite{ME11} and references therein.) 

The past fifteen years have seen great progress in applying ChPT to nuclear forces. As a result,
NN potentials of high precision have been constructed, which are based on ChPT carried to 
next-to-next-to-next-to-leading order (N$^3$LO). We will apply here the chiral NN potential 
of Ref.~\cite{EM03} which uses a cutoff, $\Lambda$, equal to 500 MeV.

A great advantage of the EFT approach to nuclear forces is that it creates two- and many-body forces
on an equal footing. 
Three-nucleon forces make their appearance at the third order in the chiral power 
counting. These leading-order contributions are: the long-range two-pion exchange graph;
the medium-range one-pion exchange diagram; and the short-range contact term.

In Ref.~\cite{Holt},  density-dependent corrections to the in-medium NN interaction have been derived from                  
the leading-order chiral 3NF. These are 
effective two-nucleon interactions that reflect the underlying three-nucleon forces and are 
therefore 
computationally very convenient, whereas realistic models of three-nucleon forces would be 
prohibitive. 

A total of six one-loop diagrams contribute at this order. Three are generated by the two-pion 
exchange graph of the chiral three-nucleon interaction and depend on the low-energy constants
$c_{1,3,4}$, which are fixed in the NN system \cite{EM03}. We use $c_1=-0.81$ GeV$^{-1}$,     
 $c_3=-3.2$ GeV$^{-1}$, and 
 $c_4=5.4$ GeV$^{-1}$.     
Two are generated by the one-pion exchange diagram and depend on the low-energy constant 
$c_D$. Finally, the short-range component depends on the constant $c_E$. 
The constants $c_D$ and $c_E$ can be fixed by fitting properties of few-nucleon systems, such as the triton
and $^3$He binding energies \cite{Nav07}. We use 
$c_D = 5.0$ and 
$c_E = 0.48$.     

In pure neutron matter, the contributions proportional to the low-energy constants $c_4,c_D$, and 
$c_E$ vanish \cite{Holt}. 
Analytical expressions for these corrections are provided in Ref.~\cite{Holt} in terms of the well-known 
non-relativistic two-body nuclear force operators. These 
can be conveniently 
incorporated in the usual NN partial wave formalism and the conventional BHF theory.

\section{Results} 
\begin{figure}[!t] 
\centering         
\vspace*{1.2cm}
\hspace*{0.5cm}
\scalebox{0.3}{\includegraphics{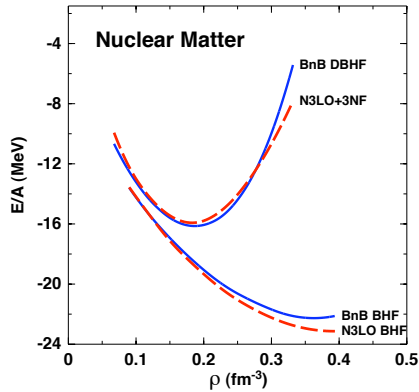}}
\vspace*{-2.0cm}
\caption{Energy/particle in SNM. The solid (blue) lines are obtained with 
Bonn B + DBHF (upper curve) and Bonn B + BHF (lower curve). The dashed (red) lines show the 
predictions by the N$^3$LO potential with (upper curve) and without (lower curve) chiral 
three-nucleon forces. 
} 
\label{snm}
\end{figure}

In Fig.~\ref{snm}, we display the EoS of SNM obtained with the two approaches outlined above.
The (blue) solid lines denote the results from the meson-exchange Bonn potential and the 
conventional BHF approximation (``BnB BHF") or the relativistic DBHF (``BnB DBHF"), while the
chiral calculations are shown by the (red) dashed lines, with and without 3NF, as denoted. 

As to be expected, both the conventional BHF calculation and the one using only the two-body chiral
potential N$^3$LO display excessive attraction and are unable to produce saturation up to 
very high density. The ``Dirac effect" brings in a powerful saturation mechanism as seen by
comparing the two (blue) solid curves. A very similar effect, both qualitatively and 
quantitatively, is generated by the effective chiral 3NF. 

In spite of the differences between the two approaches, the predictions are very close, not
only in the value of the energy at saturation, 
but also the density dependence of the saturation
mechanism.                                                          
We notice, though, that 
the DBHF energies tend to grow faster at the higher densities. 

\begin{figure}[!t] 
\centering         
\vspace*{1.2cm}
\hspace*{0.5cm}
\scalebox{0.3}{\includegraphics{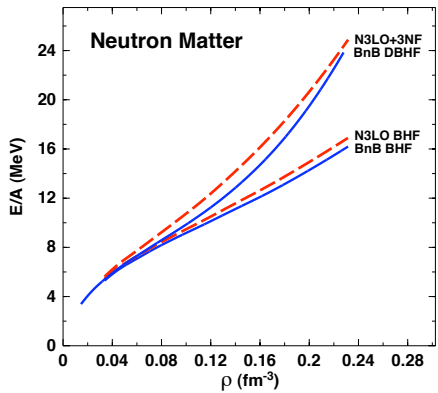}}
\vspace*{-0.5cm}
\caption{Same as Fig.~3, but for neutron matter.}                                            
\label{nm}
\end{figure}

The same comparison is done in 
Fig.~\ref{nm} for neutron matter, yielding similar conclusions.                             
Here, too, the DBHF energies show a faster growth as density increases.

In Fig.~5, we display the symmetry energy as obtained with  the 
well-known parabolic approximation, $e_{sym}=e_{NM}-e_{SNM}$. 
(Notice that 
at this point we only consider the versions of the BnB- and the N3LO-based calculations which
have realistic saturation properties, see Fig.~3.) 

The slope of the symmetry energy, usually defined in terms of the $L$ parameter, 
($L=3\rho_0 (\frac{de_{sym}}{d\rho})_{\rho_0}$), 
is a measure of the pressure gradient between neutron and symmetric matter. A variety of
experiments are aimed at constraining this important quantity \cite{Tsang+}, which correlates nearly
linearly with the neutron skin of neutron-rich nuclei.   
At the present time there seems to be a consensus from various experiments that the acceptable range
of values of the symmetry energy and its slope at saturation are centered around $e_{sym}$ and 
$L$ equal to 32.5 MeV and 70 MeV, respectively \cite{Tsang+}. 
The values of the $L$ parameter obtained with ``BnB+DBHF" and ``N3LO+3NF" are 69 and 61 MeV, respectively.
Notice that the typical uncertainty on the $L$ parameter is as large as 20-25 MeV \cite{Tsang+}.

How the differences/similarities noted above impact neutron star bulk properties is examined in Fig.~6.   
(Since our focal point is a comparison between two theoretical approaches, rather than a 
detailed calculation of $\beta$-stable matter composition, we consider only neutron 
stars made of pure neutron matter.) We also note that, in order 
to extend the chiral predictions to the high densities probed by compact stars,
we fit a three-parameter function, $e(\rho)$=$\alpha \rho + \beta \rho^{\gamma}$, to the 
``N3LO+3NF" predictions of Fig.~\ref{nm} and use this {\it ansatz} to obtain the high-density EoS needed 
for neutron star calculations. 

The comparison in Fig.~6 shows a larger star maximum mass for ``BnB DBHF", most likely the result of 
larger repulsion of that model at high density. 
Overall, the predicted star bulk properties are 
only moderately different in the two approaches.

\begin{figure}[!t] 
\centering         
\vspace*{1.2cm}
\hspace*{0.5cm}
\scalebox{0.3}{\includegraphics{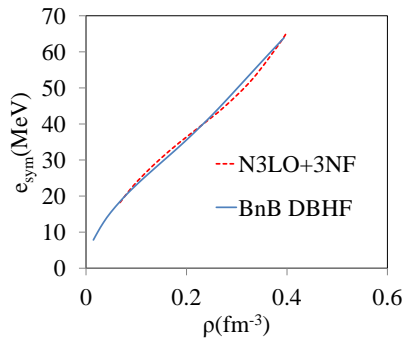}}
\vspace*{-3.0cm}
\caption{The symmetry energy {\it vs.} density for the models
corresponding to the two upper curves in Fig.~3-4.} 
\label{5}
\end{figure}

\begin{figure}[!t] 
\centering         
\vspace*{3.2cm}
\hspace*{0.5cm}
\scalebox{0.3}{\includegraphics{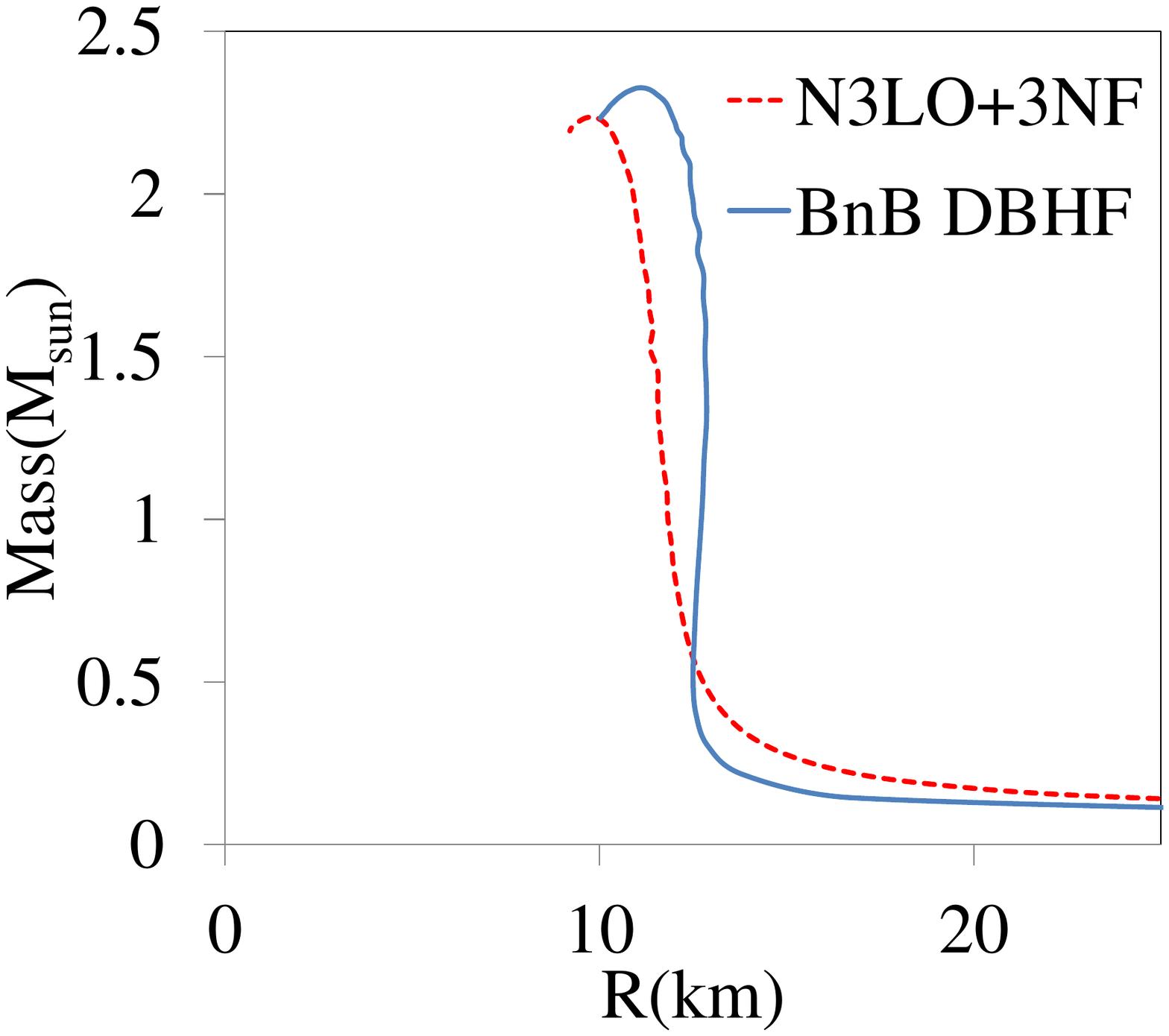}}
\vspace*{-3.0cm}
\caption{Neutron star mass-radius relation as predicted by the two models                             
as denoted.} 
\label{6}
\end{figure}

\section{Conclusions}                                                                  
We have considered two very different methods                                           
to approach the study of nucleonic matter: 
one based on a meson-theoretic 
potential and the Dirac-Brueckner-Hartree-Fock approximation; the other based on a high-precision
chiral NN potential and chiral effective three-nucleon forces at NNLO. 
The predictions we have considered include: the EoS of nuclear and neutron matter, the
symmetry energy and its slope, and the mass-radius relation in a neutron star. 

From our results, 
we conclude that the DBHF method is an excellent phenomenology capable of incorporating 
important many-body effects that are crucial to nuclear saturation.

In both approaches, the effective 3NF is generated by one nucleon interacting with the Fermi sea.
That is, in both cases we have 
effective two-nucleon interactions that reflect the underlying three-nucleon forces.            
No matter if this interaction proceeds {\it via} relativistic meson exchange or {\it via} chiral EFT
forces, the results are very similar. This is reassuring and confirms that the two ways of
describing nuclear forces are complementary.

\section*{Acknowledgments}
Support from the U.S. Department of Energy under Grant No. DE-FG02-03ER41270 is 
acknowledged, and from                                                                  
 the Italian Ministero dell'Istruzione dell'Universit\'a della Ricerca (MIUR) under
PRIN 2009. 
We are grateful to F. Weber for providing his TOV code for the calculation of neutron 
star properties. 

\end{document}